\documentstyle[12pt]{article}
\begin{document}
\baselineskip 22pt

\begin{center}
 {\Large HOW UNCONDITIONALLY SECURE QUANTUM \\ 
BIT COMMITMENT IS POSSIBLE} \\
\vspace*{.4in}

{\Large Horace P. Yuen} \\  {\large Department of Electrical and
Computer Engineering \\ Department of Physics and Astronomy\\
Northwestern University \\ Evanston IL  60208-3118 \\ email:
yuen@ece.northwestern.edu}
\end{center}
\vspace*{.4in}

\begin{abstract}
Bit commitment involves the submission of evidence from one party to
another so that the evidence can be used to confirm a later revealed
bit value by the first party, while the second party cannot determine
the bit value from the evidence alone. It is widely believed that
unconditionally secure quantum bit commitment is impossible due to
quantum entanglement cheating, which is codified in a general
impossibility theorem. In this paper, the scope of this general
impossibility proof is analyzed, and gaps are found.  Two
variants of a bit commitment scheme utilizing anonymous
quantum states and decoy states are presented. In the first variant, the exact verifying measurement is
independent of the committed bit value, thus the second party can make
it before the first party opens, making possible an unconditional security proof based on no-cloning. In the second variant, the impossibility proof fails because quantum entanglement purification of a mixed state does not render the protocol determinate. Whether impossibility holds in this or similar protocols is an open question, although preliminary results already show that the impossibility proof cannot work as it stands.

\vspace*{.2in}

\noindent PACS \#: 03.67Dd, 03.65Bz
\end{abstract}

\vspace*{.5in}

Note:

We have made a few clarifications and elaborations in this revision.

\newpage

\renewcommand{\thesection}{\Roman{section}}

\tableofcontents

\newpage

\section{\hspace{.2in}Introduction}

\indent 
Quantum cryptography \cite{bennett}, the study of information security
systems involving quantum effects, has recently been associated almost
exclusively with the cryptographic objective of key distribution.
This is due primarily to the nearly universal acceptance of the general
impossibility of secure quantum bit commitment (QBC), taken to be a
consequence of the Einstein-Podolsky-Rosen (EPR) type entanglement
cheating which rules out QBC and other quantum protocols that have
been proposed for various other cryptographic objectives
\cite{brassard}.  In a bit commitment scheme, one party, Adam,
provides another party, Babe, with a piece of evidence that he has
chosen a bit b (0 or 1) which is committed to her.  Later, Adam would
``open'' the commitment: revealing the bit b to Babe and convincing
her that it is indeed the committed bit with the evidence in her
possession.  The usual concrete example is for Adam to write down the
bit on a piece of paper which is then locked in a safe to be given to
Babe, while keeping for himself the safe key that can be presented
later to open the commitment.  The evidence should be {\em binding},
i.e., Adam should not be able to change it, and hence the bit, after
it is given to Babe.  It should also be {\em concealing}, i.e., Babe
should not be able to tell from it what the bit b is.  Otherwise,
either Adam or Babe would be able to cheat successfully.

In standard cryptography, secure bit commitment is to be achieved
either through a trusted third party or by invoking an unproved
assumption on the complexity of certain computational problem.  By
utilizing quantum effects, various QBC schemes not involving a third
party have been proposed that were supposed to be unconditionally
secure, in the sense that neither Adam nor Babe can cheat with any
significant probability of success as a matter of physical laws.  In
1995-1996, a general proof on the impossibility of unconditionally
secure QBC and the insecurity of previously proposed protocols were
described \cite{mayers1}-\cite{lo}.  Henceforth, it has been generally
accepted
that secure QBC and related objectives are impossible as a matter of
principle \cite{lo1}-\cite{lo4}.

There is basically just one impossibility proof, which gives the EPR
attacks for the cases of equal and unequal density operators that Babe
has for the two different bit values. The proof shows that if Babe's
successful cheating probability $P^B_c$ is close to the value 1/2,
which is obtainable from pure
guessing of the bit value, then Adam's successful cheating probability $P^A_c$ is
close to the perfect value 1.  This result is stronger than the mere impossibility of unconditional security, namely that it is
impossible to have both $P^B_c \sim 1/2$ and $P^A_c \sim 0$. Since
there is no known characterization of all possible QBC protocols,
logically there can really be no general impossibility proof even if
it were indeed impossible to have an unconditionally secure QBC
protocol. 

In this paper, the formulation within which the general impossibility
proof was developed will be analyzed.  The mechanism for the success
of the impossibility proof within a limited scope will be delineated.
It is shown that the use of classical randomness unknown to one of the
two parties, common in many standard cryptographic protocols, is not
properly accounted for in the previous impossibility proof
formulation.  In particular, the turning of classical randomness into
quantum determinateners via quantum purification of a mixed quantum
state does {\em not} render a quantum protocol determinate with no
further role for classical randomness, as described in the
impossibility proof.  Specifically, a scheme utilizing anonymous
states and decoy states will be presented, and the different ways in which the impossibility proof fails for
these variants will be explicitly pinpointed. The results are
developed within nonrelativistic quantum mechanics, unrelated to relativistic
protocols \cite{kent} or cheat-sensitive protocols \cite{hardy}. 
Since bit commitment leads to ``coin-tossing'' and other cryptographic protocols, our present results have
immediate impact on many recent works on quantum coin-tossing and multiparty computation.

To provide a foretaste of the failure of the impossibility proof, the
following two points may be mentioned. First, the impossibility proof
has {\em no} role for any possible classical randomness that Babe may
introduce, which, even after quantum purification, would actually be
explicitly used by her in her verification of the bit.  If the use of
such randomness by Babe is taken into account, it is not hard to see
that the success of Adam's EPR cheat may depend on knowing the actual
value of such random numbers.  Secondly, there are concealing
protocols for which Babe can make all the measurements for
verification {\em before} Adam opens because the verifying measurement is independent of the
bit value, with no consequent possibility that an 
information carrying state needs to be discarded due to measurement
basis mismatch.  This kind of protocol is one of several types outside the impossibility proof formulation.  Indeed, a general formulation of all possible QBC protocols is not yet available that includes a proper expression of just the concealing condition, not to mention both concealing and binding with corresponding expressions for the cheating probabilities.

In section II, the impossibility proof would be described and
extended.  The mechanism of its success within its limited scope will
be highlighted.  In section III, the use of anonymous states in QBC
will be developed, in which Babe uses classical random numbers in the
most direct way in protocols involving two-way quantum communication.
It is explicity demonstrated that the impossibility proof, 
specifically the use of the doctrine ``Church of Larger Hilbert 
Space,'' fails to cover such situations in two different ways.  In section IV, our basic scheme is introduced in a
preliminary form which is not yet unconditionally secure but which
already invalidates the impossibility proof.  Two variants of
the scheme are described.  One of which, QBCp3m, allows Babe to make
perfect verifying measurements before Adam opens.  The reader is urged to first 
read Appendix D for a concise presentation of this basically rather 
simple protocol, as it confirms our statement
above that there can be no general impossibility proof without a
characterization of all possible QBC protocols.  In section V, the
protocol QBCp3m is extended to fully unconditionally secure
ones together with their security proofs.  Some general and practical
observations are made in the last section VI. Note that the same 
index may denote different quantities in different sections, and 
the notation $ \otimes$ is often omitted for brevity.

\newpage

\section{\hspace{.2in}The Impossibility Proof}

\indent
In this Section we review the standard formulation of the
impossibility proof, present some pertinent new results, and explain the
precise mechanism of the EPR cheating.

According to the impossibility proof, Adam would generate $|\Phi _0
\rangle$ or $|\Phi _1 \rangle$ depending on b = 0 or 1,
\begin{equation}
|\Phi _0 \rangle = \sum_i \sqrt{p_i} | e_i \rangle | \phi _i \rangle,
\label{ent1}
\end{equation}
\begin{equation}
 |\Phi _1 \rangle = \sum_i \sqrt{p'_i} | e'_i \rangle | \phi ' _i
  \rangle
\label{ent2}
\end{equation}
where the states $\{ | \phi_i \rangle \}$ and $\{ | \phi'_i \rangle
\}$ in ${\mathcal H}^B$ are openly known, $i \in \{1, \ldots, M \}$,
$\{ p_i \}$ and $\{ p'_i \}$ are known probabilities, while $\{ | e_i
\rangle \}$ and $\{ | e'_i \rangle \}$ are two complete orthonormal
sets in ${\mathcal H}^A$. All Dirac kets are normalized in this
paper. Adam sends Babe ${\mathcal H}^B$ while keeping ${\mathcal H}^A$
to himself.  He opens by measuring the basis $\{ | e_i \rangle \}$ or $\{
| e'_i \rangle \}$ in $\mathcal{H}^A$ according to his committed state
$| \Phi_0 \rangle$ or $| \Phi_1 \rangle$, resulting in a specific $|
\phi_i \rangle$ or $| {\phi'}_i \rangle$ on $\mathcal{H}^B$, and
telling Babe which $i$ he has obtained.  Babe verifies by measuring
the corresponding projector and will obtain the value 1 (yes) with
probability 1. In this formulation, Adam can switch between $|\Phi _0
\rangle$ and $|\Phi _1 \rangle$ by operation on ${\mathcal H}^A$
alone, and thus alter the evidence to suit his choice of b before
opening the commitment.  In the case $\rho^B_0 \equiv {\rm tr}_A |\Phi
_0 \rangle \langle \Phi _0 | = \rho ^B _1 \equiv {\rm tr}_A |\Phi _1
\rangle \langle \Phi _1|$, the switching operation is to be obtained
by using the so-called ``Schmidt decomposition
\cite{schmidt_decomp},'' the expansion of $|\Phi _0 \rangle$ and
$|\Phi _1 \rangle$ in terms of the eigenstates $|\hat{\phi}_k \rangle$
of $\rho^B_0 = \rho^B_1$ with eigenvalues $\lambda_k$ and the eigenstates $|\hat{e}_k \rangle$ and
$|\hat{e}'_k \rangle$ of $\rho^A_0$ and $\rho^A_1$,
\begin{equation}
|\Phi _0 \rangle = \sum _k \sqrt{\lambda_k} |\hat{e}_k \rangle
|\hat{\phi}_k \rangle, \hspace*{.2in} |\Phi _1 \rangle = \sum _k
\sqrt{\lambda_k} |\hat{e}'_k \rangle |\hat{\phi}_k \rangle
\end{equation}
By applying a unitary $U^A$ that brings $\{ |\hat{e}_k \rangle \}$ to
$\{ |\hat{e}'_k \rangle \}$, Adam can select between $|\Phi_0 \rangle$
or $|\Phi_1 \rangle$ any time before he opens the commitment but after
he supposedly commits.  When $\rho_0^B$ and $\rho_1^B$ are not equal
but close, it was shown that one may transform $| \Phi _0 \rangle$ by
an $U^A$ to a $| \tilde{\Phi} _0 \rangle$ with $|\langle \Phi _1 |
\tilde{\Phi} _0 \rangle |$ as close to 1 as $\rho ^B_0$ is close to
$\rho ^B_1$ according to the fidelity F chosen, and thus the state $|
\tilde{\Phi} _0 \rangle$ would serve as the effective EPR cheat.

In addition to the above quantitative relations, the gist of the
impossibility proof is supposed to lie in its generality -- that any
QBC protocol could be fitted into its formulation, as a consequence of
various arguments advanced in \cite{mayers1}-\cite{lo4}. Among
other reasons, it appeared to the
present author from his development of a new cryptographic tool,
anonymous quantum key technique \cite{yuen_capri}, that the
impossibility proof is not sufficiently general. Since there
is no need for Adam to entangle anything in an honest protocol.
Adam can just send Babe a state $| \phi_i \rangle$
with probability $p_i$ when he picks b=0.  When he picks b=1, he sends $| \phi'_i
\rangle$ with probability $p'_i$.  If the anonymous key technique is
employed, $| \phi_i \rangle$ and $| \phi'_i \rangle$ are to be
obtained from applying $U_{0i}$ or $U_{1i}$ from some fixed openly known
set of unitary operators $\{ U_{0i} \}$ and $\{ U_{1i} \}$ on ${\mathcal
H}^B$ by Adam to the states $| \psi \rangle$ sent to him by Babe and
known only to her. As a consequence, Adam would not be able to
determine the cheating unitary transformation $U^A$.  This use of anonymous states is {\em not}
explicitly accounted for in the open literature, and the role of
classical random numbers in the problem formulation is not clearly and
fully laid out in the impossibility proof.  However, it seems the
prevailing opinion is that the impossibility proof covers classical
randomness in essence, basically through the use of quantum
purification of classical
randomness \cite{mayers}, \cite{lo3}, \cite{miller-quade}.  This claim that the impossibility proof covers all
classical randomness has never been explicitly demonstrated, and it is
one major purpose of  this paper to show that such a claim is
erroneous.  The gap in the reasoning, to be delineated in section
III, is best appreciated after a careful quantitative development of
the impossibility proof to be presently given.

In a QBC protocol, the $\{ | \phi_i \rangle \}$ and $\{ | \phi'_i \rangle \}$
are chosen so that they are concealing as evidence, i.e. Babe cannot
reliably distinguish them in optimum binary hypothesis testing
\cite{helstrom}. They would also be binding if Adam is honest and sends them as they
are above, which he could not change after Babe receives them.  Babe
can always guess the bit with a probability of success $P^B_c = 1/2$,
while Adam should not be able to change a committed bit at all.
However, it is meaningful and common to grant {\em unconditional
security} when the best $\bar{P}^B_c$ Babe can achieve is arbitrarily
close to 1/2 and Adam's best probability of successfully changing a
committed bit $\bar{P}^A_c$ is arbitrarily close to zero even when both parties have perfect technology and unlimited resources
including unlimited computational power \cite{mayers}.  

The operation of unitary transformation with subsequent measurement of
an orthonormal basis is equivalent to the mere measurement of another
orthonormal basis $\{ |\tilde{e}_i \rangle \}$ on the system.  Thus,
the net cheating operation can be described by writing
\begin{equation}
| \Phi_0 \rangle = \sum_i \sqrt{\tilde{p}_i} | \tilde{e}_i \rangle |
  \tilde{\phi}_i \rangle,
\end{equation}
\begin{equation}
\sqrt{\tilde{p}_i}|\tilde{\phi}_i \rangle \equiv \sum_j \sqrt{p_j}
V_{ji} |\phi_j \rangle
\end{equation}
for a unitary matrix V defined by $| e_i \rangle = \sum_j V_{ij} |
\tilde{e}_j \rangle$, and then measuring $|\tilde{e}_i\rangle$. For
convenience, we may still in the rest of the paper refer to the
cheating operation as a $U^A$ transformation described at the
beginning of this Section, with  $| e_i \rangle =  U^A
|\tilde{e}_i \rangle$.  From (5), the $| \tilde{\phi}_i \rangle$ obtainable by operation on
${\mathcal H}^A$ alone are some unitary linear combinations of the $|
\phi_i \rangle$. The quantitative expression
for $P^A_c$ can 
now be given.  If Babe verifies the individual $| \phi'_i \rangle$,
the Adam's successful cheating probability is
\begin{equation}
P^A_c = \sum_i \tilde{p}_i | \langle \tilde{\phi}_i | \phi'_i \rangle
|^2.
\end{equation}
When randomness from Babe is present, further averaging is needed to yield the final $P^A_c$. The EPR cheating mechanism is clear from (5)---via entanglement and
measurement of a different basis, Adam can generate {\em unitary linear
combinations of the committed states} $|\phi_i \rangle $ to
approximate the states  $|\phi'_i \rangle $.  The approximation is
guaranteed to be good when the protocol is concealing, as follows.

In general, the optimal cheating probability $\bar{P}^B_c$ for Babe is
given by the probability of correct decision for optimally discriminating between two density operators $\rho^B
_0$ and $\rho^B _1$ by any quantum measurement. For equal a priori probabilities,

\begin{equation}
\bar{P}^B_c = \frac{1}{4} (2 + \| \rho^B _0 - \rho^B _1 \| _1)
\end{equation}
where $\| \cdot \| _1$ is the trace norm, $\| \tau \| _1 \equiv
tr(\tau ^{\dagger} \tau )^{1/2}$, for a trace-class operator $\tau$
\cite{schatten}.  In terms of a security parameter $n$ that can be
made arbitrarily large, the statement of unconditional security (US) can be
quantitatively expressed as
\begin{equation}
{\rm (US)} \quad \qquad \lim _n \bar{P} ^B _c = \frac {1} {2} \quad {\rm
and} \quad \lim _n \bar{P} ^A _c = 0.
\end{equation}
Condition (US) is equivalent to the statement that for any $\epsilon > 0$, there exists an
$n_0$ such that for all $n > n_0$, $\bar{P}^B_c - \frac{1}{2} <
\epsilon$ and $\bar{P}^A_c < \epsilon$, i.e. $\bar{P}^B_c -
\frac{1}{2}$ and $\bar{P}^A_c$ can both be made arbitrarily small for
sufficiently large $n$.  The impossibility proof claims more
than the mere impossibility of (US), it asserts \cite{mayers1} the
following statement (IP):
\begin{equation}
{\rm (IP)} \qquad\bar{P}^B_c = \frac{1}{2} + O(\frac{1}{n}) \Rightarrow \bar{P}^A_c = 1
- O(\frac{1}{n})
\end{equation}
Condition (9) implies the following limiting statement
\begin{equation}
{\rm (IP')} \qquad \lim _n \bar{P} ^B _c = \frac {1} {2} \quad
\Rightarrow \quad \lim _n \bar{P} ^A _c = 1.
\end{equation}
that directly contradicts (8).  One may regard (IP') as the general
impossibility statement, independently of the specific convergence rate of (9). In the $\rho ^B _0 = \rho ^B _1$ case, the EPR cheat shows that
$\bar{P}^B_c=\frac{1}{2}$ implies $\bar{P}^A_c=1$.  Thus (IP')
generalizes it to the assertion that the function
$\bar{P}^A_c(\bar{P}^B_c)$, obtained by varying $n$, is {\em
continuous} from above at $\bar{P}^B_c = \frac{1}{2}$. Note the
difference between the truth of (IP') and the weaker statement that (US) is impossible. In the middle ground
that $\lim_n \bar{P}^B_c = \frac{1}{2}$ implies just $0 < \lim_n
\bar{P}^A_c < 1$, the protocol would be concealing for Babe and
quantitatively cheat-sensitive for Adam.  However, it may be
expected that if $\bar{P}^A_c $ is not close to $1$, it may be made close
to $0$ in an extension protocol which thus becomes unconditionally
secure. 

The cheating transformation for the $\rho^B_0 \neq \rho^B_1$ case is
determined from ref. \cite{jozsa} according to the impossibility proof [3]-[4], which would
proceed as follows. Let $|\lambda_i\rangle$ and $|\mu_i\rangle$ be the eigenstates of
$\rho^B_0$ and $\rho^B_1$ with eigenvalues $\lambda_i$ and $\mu_i$.
The Schmidt normal forms of the purifications $|\Phi_0\rangle$ and $|
\Phi_1 \rangle$ of $\rho^B_0$ and $\rho^B_1$ are given by
\begin{equation}
|\Phi_0\rangle = \sum_i \sqrt{\lambda_i} |f_i\rangle |\lambda_i\rangle,
\end{equation}
\begin{equation}
|\Phi_1\rangle = \sum_i \sqrt{\mu_i} |g_i\rangle |\mu_i\rangle
\end{equation}
for complete orthonormal sets $\{|f_i\rangle\}$ and $\{|g_i\rangle\}$
on ${\mathcal H}^A$.  Define the unitary operators $U_0$, $U_1$ and
$U_2$ by
\begin{equation}
U_0|\lambda_i\rangle = |\mu_i\rangle,
\end{equation}
\begin{equation}
U_1|\lambda_i\rangle = |f_i \rangle,
\end{equation}
\begin{equation}
U_2|\mu_i\rangle=|g_i \rangle.
\end{equation}
Let $U$ be the unitary operator for the polar decomposition of
$\sqrt{\rho^B_0} \sqrt{\rho^B_1}$ ,
\begin{equation}
\sqrt{\rho^B_0} \sqrt
{\rho^
B_1} = \left| \sqrt{\rho^B_0}
\sqrt{\rho^B_1} \right| U.
\end{equation}
Then $\left| \langle \Phi_0 | \Phi_1 \rangle \right|$ assumes its
maximum value $F(\rho^B_0,\rho^B_1), F(\rho_0,\rho_1) \equiv {\rm
tr} \sqrt{\sqrt{\rho_0} \rho_1 \sqrt{\rho_0}}$, when
\begin{equation}
U U^T_2 U_0 U_0^T U^{T^\dagger}_1 = I
\end{equation}
where $T$ denotes the transpose operation. Thus, when $\rho^B_0$, $\rho^B_1$, and $|e_i\rangle$ are given,
$|g_i \rangle = |e'_i\rangle$ of $|\Phi_1\rangle$ is determined from (12) via solving for $U_2$
from (17).  In general, these $U$'s are isometries, but the above relations still hold.

The above formulation (11)-(17), utilizing Jozsa's proof \cite{jozsa} of
Uhlmann's theorem, covers both the $\rho^B_0 = \rho^B_1$ case and the
$U^A = I$ (i.e., $|\Phi_0 \rangle = | \Phi_1 \rangle$) situation as
special cases.  Apparently form these equations, knowledge of the
eigenstates of $\rho^B_0$ and  $\rho^B_1$  is required to find the
cheating transformation $U^A$ that brings $|e_i \rangle$ to $| \tilde{e}_i
\rangle$. Actually, both (11)-(17) and the Schmitt  decomposition  obscure the
underlying mechanism of the EPR cheating given by (5).  In the present
context, they suggest that knowledge of the $\rho^B_b$ eigenstates is
needed to determine $U^A$, which is actually much simpler determined
by the following

\vspace{3ex}
\noindent
{\em Theorem} 2:

\indent
The $U^A$ that maximizes $|\langle \tilde{\Phi}_0 | \Phi_1 \rangle |$,
defined through the matrix ${\bf U}, U_{ij} \equiv \langle e_i | U^A
|e_j \rangle$, is determined by

\begin{equation}
{\bf \Lambda U} = | {\bf{\Lambda}} |
\end{equation}
where
 
\begin{equation}
{\Lambda}_{ij} \equiv {\sqrt{p'_i p_j}}
\langle \phi'_i | \phi_j \rangle, \quad 
|{\bf{\Lambda}}| \equiv ({\bf{\Lambda \Lambda^\dagger}}
)^{\frac12}
\end{equation}
When $p_i = p'_i$, the corresponding 
\begin{equation} 
\tilde{P}^{A}_{c} = \sum_i
\left( |{\bf{\Lambda}}|_{ii} \right)^2
\end{equation}
which satisfies
\begin{equation}
F^2 \leq \tilde{P}^A_c \leq F
\end{equation}
\noindent
The lower bound in (21) is valid also for $p_i \neq p'_i$.

\vspace{3ex}
Theorem 2 is proved in Appendix A.  Note that in terms of the
${\bf V}$ in (5), ${\bf U} = {\bf V}^T$. The
bounds (21) simply characterize $\tilde{P}^{A}_{c}$ in terms of $F$, and yield 
$\bar{P}^A_c \geq F^2$ for the actual optimal probability $\bar{P}^A_c$ that 
maximizes (6).  This lower bound yields the usual impossibility proof \cite{mayers1} or (IP)
of (9) when combined with the lower bound on $|| \cdot ||_1$ in terms of $F$ \cite{fuchs}.  
When ${\bf \Lambda}$ is invertible, ${\bf U} = {\bf \Lambda}^{-1}
|{\bf \Lambda}|$ from (18).   In general, one
does not need to compute the eigenstates of $\rho^B_b$ to find $U^A$,
which is determined through ${\bf \Lambda}$ that is given directly in
terms of the known states and probabilities.

\newpage

\section{\hspace{0.2in}The Impossibility Proof and Anonymous
States}

\indent
The use of anonymous states by Babe as briefly described in the last
section is just one obvious way to introduce classical randomness for
her in a QBC protocol, which appears to thwart Adam's EPR cheating by
denying him the knowledge to find the proper cheating transformation.
The ways in which the impossibility proof fails in this situation are
detailed in this section.

In general, such use of anonymous states by Babe can be described as
follows. She sends Adam a state $|\psi \rangle \in \mathcal{H}^B$ known only to
herself. Depending on b = 0 or 1, Adam applied to $|\psi \rangle$ a unitary
operator $U_{bi} , i \in \{1; \ldots , M \}$  with probabilities $p_i$ or $p'_i$.  In the notation
of section II,
\begin{equation}
| \phi_i \rangle = U_{0i} | \psi \rangle,  \quad |\phi'_i \rangle = U_{1i} |
\psi \rangle
\end{equation}

\noindent
Adam sends the modulated state back to Babe, and opens by revealing
b and $i$.  He can form the entangled $ | \Phi_0 \rangle$ by
applying the unitary operator $U_0$ on $\mathcal{H}^A \otimes
\mathcal{H}^B$,
\begin{equation}
U_0 = \sum_i |e_i \rangle \langle e_i | \otimes U_{0i}
\end{equation}
with initial state $|A \rangle \in \cal{H}^A$ satisfying
$\langle e_i | A \rangle = \sqrt{p_i}$.  It appears
from Theorem 2 above that the cheating transformation $U^A$ as
determined by 
$\langle \phi'_{i} |
\phi_j \rangle = \langle \psi | U^{\dagger}_{1i} U_{0j} | \psi
\rangle$ would depend on $|\psi \rangle$ in general, thus cannot
be found by Adam.  The impossibility proof handles this situation rather explicitly in
\cite{mayers}, \cite{lo1}, \cite{lo4}, and \cite{miller-quade}, in the following way.

The state $|\psi \rangle$ is supposed to be picked by Babe from a set $\{| \psi_k \rangle\}, k \in \{ 1, \ldots, L \}$ with
probabilities $\lambda_k$ that are all openly known.  The associated classical
randomness is then purified by having Babe generate the entangled
state
\begin{equation}\label{label.B} \label{eq23}
| \Psi \rangle = \sum_k \sqrt{\lambda_k} | \psi_k \rangle | f_k \rangle,
\end{equation}
where the $|f_k \rangle$'s are complete orthonormal in
$\mathcal{H}^C$, send  $\mathcal{H}^B$ to Adam
while keeping  $\mathcal{H}^C$ to herself.  At the end of the commitment phase she
would measure $\{ | f_k \rangle \}$ to pin down a specific $| \psi_k \rangle$.  The proof, however,
is {\em not} carried to the end, and the above description is
considered sufficient to ensure that the impossibility proof works in
the presence of classical randomness introduced by Babe---from
quantum entanglement purification of a mixed state and postponement of
all measurements to end of commitment, the classical randomness is
rendered quantum-mechanically determinate and everything is known to
Adam again for him to find the cheating transformation $U^A$.
While Babe may not actually form $| \Psi \rangle$, the so-called ``Church of Larger Hilbert Space'' doctrine \cite{gottesman} is used to justify the equivalence.
In the following, it will be shown that the equivalence does not hold when Babe does something else to cheat, and that theimpossibility proof does not go through even when Babe actually forms
$|\Psi \rangle$ and postpone her measurement on $\mathcal{H}^C$ until after
Adam opens.

To spell out the impossibility proof argument, one actually needs to show that $U^A$
is independent of $\{ |f_k \rangle \}$ in $\mathcal{H}^C$ and
Adam only needs $\mathcal{H}^B$, not the full $| \Psi \rangle$ of (24), to form his entanglement.  These turn out to be true as a
consequence of (18) in Theorem 2 above.  However, it is against common
probabilistic intuition that randomness would altogether disappear (to
Adam) upon a quantum interpretation.  There is no reason why Babe has
to generate (24) instead of any specific $|\psi_k \rangle$.  More generally, to
form $| \Psi \rangle$, Babe can choose any probability
distribution on $\{| \psi_k \rangle \}$, not
the $\{ \lambda_k \}$ that Adam believes. It is not a meaningful formulation to assume that Adam knows $\{ \lambda_k \}$. Thus, one {\em cannot} eliminate via quantum
purification what is nonrandom to Babe (upon her choice or measurement) and random to
Adam \cite{adam1}.  Indeed, Babe can generate any state $| \psi \rangle \in \mathcal{H}^B$, or any entangled state $| \Phi \rangle \in {\cal H}^B \otimes {\cal H}^C$ for any ${\cal H}^C$ that she keeps to herself. A
careful formulation for the concealing condition needs to be
developed.  

To show the inadequacy of the formulation of the impossibility proof,
assume that $\rho^B_0 (\Psi)$ and $\rho^{B}_{1} (\Psi)$ are indeed close from the
use of (22) and (24) with $\lambda_k = 1/L$ for $L$ large.
Let $|\psi_1 \rangle$ be such that $\rho^B_0 (\psi_1 )$ and $\rho^B_1
(\psi_1 )$ are far apart, being possible even though $\rho^B_0 (\Psi)$
and $\rho^B_1 (\Psi)$ are
close because $1/L$ is small. Then Babe can cheat by using
$\lambda_1 = 1$ instead of $\lambda_k = 1/L$.  To ensure a
concealing protocol, one must impose the uniformity condition
\begin{equation}
\rho^B_0 (\psi) \approx \rho^B_1 (\psi), \quad \forall \psi \in{\mathcal{H}}^B,
\end{equation}
or, more generally,
\begin{equation}
\rho^{BC}_0 (\Psi) \approx \rho^{BC}_1(\Psi),\quad \forall \Psi \in {\cal H}^B \otimes {\cal H}^C
\end{equation}
for any ${\cal H}^C$ Babe can use to entangle, with $\approx$ being taken in the sense of trace norm from
(7).  Such a concealing condition has not been given in the literature, 
but it is needed whenever a state is passed from Babe to Adam 
in a proper formulation of the problem. From the condition $\rho^{BC}_0
(\Psi ) \approx \rho^{BC}_1 (\Psi )$ for a fixed $\Psi$, one may at best conclude that
$\rho^B_0 (\psi_k ) \approx \rho^B_1 (\psi_k )$ for those $k$ where
$\lambda_k \not\approx 0$. Thus, the impossibility proof {\em errs} in
asserting that Adam can cheat under the above condition---he cannot,
and Babe can instead.  In Appendix B, an example is given in which
$\rho^{BC}_0(\Psi) = \rho^{BC}_1(\Psi)$ for a given $\Psi$, but Adam cannot cheat even when no $\lambda_k$ is small.

Assuming that condition (25) is satisfied, let us examine how Adam's
EPR cheating works.  If one follows the impossibility proof, it would
work if Babe verifies on the state $|\Psi \rangle$ of (24), i.e., depending on
b = 0 or 1 she checks whether $|\Psi \rangle$ becomes
\begin{equation}
U_{bi}| \Psi \rangle = \sum_k \sqrt{\lambda_k} U_{bi} |
\psi_k \rangle | f_k \rangle
\end{equation}
for the $i$ opened by Adam.  However, that is {\em not} the way she verifies
according to the protocol.  She would make a preliminary measurement of $\{| f_k
\rangle \}$ first
with result $j$ and then check whether the state is $U_{bi} |
\psi_j \rangle$. She can in
fact postpone her measurement on $\mathcal{H}^C$ until after Adam opens.  The
important point is that she is going to make a measurement and use the
result in the verification. While such measurement does not allow her to cheat any better, it 
may help defeat Adam's EPR cheating.  
In the impossibility proof there is {\em no} role given to any
classical randomness other than $\{ p_i \}$ 
and $\{ p'_i \}$---it is implicitly assumed that Babe's random
number known only to herself is {\em not} used in her verification as just described.
Such lack of utilization of possible classical randomness represents
a huge gap in the impossibility proof, making it severely limited in
scope and {\em incorrect} as a general proof. The doctrine of the
``Church of Larger Hilbert Space'' is {\em irrelevant} to the protocol
behavior as it should be; it would not make the protocol determinate. It is clear that
classical random numbers can be generated by both Adam and Babe in a
general quantum protocol, which are kept secret from the other party
and used in an essential way as in many standard cryptographic
protocols. The impossibility proof does not begin to incorporate such
possibilities. 

We examine more exactly how the impossibility proof fails in the
present situation.  The cheating transformation $U^A$ is taken
to be the one that maximizes $|\langle \Phi_1 | U^A |
\Phi_0 \rangle |$, not the {\hspace{.25in}one that maximizes $P^A_c$
of (6). However, in addition to the
lower bound in (21) that applies also to ${\bar P}^A_c$,
in general $U^A$ is determined by the inner product matrix $\langle
\phi'_i | \phi_j \rangle$, apart from the a priori probabilities.  Any
such cheating {\bf U} is thus determined via $\sum_k \lambda_k \langle
\psi_k | U^{\dagger}_{1i} U_{0j} | \psi_k \rangle \equiv
{\bar{u}}_{ij}$ for $|\Psi \rangle$,
but via $\langle \psi_k | U^{\dagger}_{1i}U_{0j} | \psi_k \rangle
\equiv u^k_{ij}$ for
each $|\psi_k \rangle$. There is no reason to
expect that the {\bf U} as determined by ${\bar{u}}_{ij}$ would be
close to the {\bf U} determined by
$u^k_{ij}$, whatever the $\lambda_k$'s are. 

Actually, one does not
need to transform among the maximizing  {\bf U} in order for
impossibility to hold.  The general problem can be cast as follows. From (22) and (5), we have
dependence of the committed and cheating states on the anonymous
state $|\psi \rangle$, to be simply denoted by $\phi_i (\psi )$, $\phi'_i
(\psi )$, and $\tilde{\phi_i} (\psi )$ for notational simplicity
dropping the Dirac kets, as already done occasionally above.  In the present formulation with only
anonymous states from Babe in the form (22), all of Adam's possible attacks are described by local
measurements and announcing a different b.  For this attack to
succeed with $\bar{P}^A_c \sim 1$ given $|\Phi_0 \rangle$ is
committed, one presumably wants 
\begin{equation}
\tilde{\phi}_i(\psi) \approx \phi'_i (\psi) , \quad \forall
\psi \in \mathcal{H}^B
\end{equation}
or
\begin{equation}
\tilde{\phi}_i(\Psi) \approx \phi'_i(\Psi), \quad \forall \Psi \in {\cal H}^B \otimes {\cal H}^C
\end{equation}
for some fixed $U^A$ or {\bf V} independent of $\psi$, the $\approx$ in (28)-(29) taken in the sense of state inner product.
Condition (28) expresses the requirement that as the anonymous state
$|\psi \rangle$ changes, the approximate state $\tilde{\phi_i}(\psi )$
must follow the b = 1 states  $\phi'_i(\psi )$.  
Strictly, the condition is only  that there exists a {\bf V} 
such that the $P^A_c \left( \psi \right)$ given by (6) satisfies
\begin{equation}
P^A_c (\psi) \approx 1 \quad , \quad \forall 
\psi \in \mathcal{H}^B
\end{equation}
where the $\psi$-dependence enters through $\phi_i (\psi)$ and $\phi'_i (\psi)$. 
The problem of impossibility (IP') becomes whether (30) holds when (25) is
satisfied.  A similar condition is obtained for $\Psi$ that includes Babe's possible entanglement of the anonymous state in ${\cal H}^B$.

In the case of perfect security, the above use of anonymous states
cannot prevent the success of EPR attacks due to

\vspace{3ex}
\noindent
{\em Theorem} 3 \cite{quick}:

\indent
\vspace{4ex}
The condition
\begin{equation}
\rho^B_0 (\psi) = \rho^B_1(\psi) \;\;\;\;, \quad \forall \psi \in \mathcal{H}^B
\end{equation}
implies, for every $i$,
\begin{equation}
\tilde{\phi}_i (\psi) = \phi'_i (\psi) \;\;\;\;, \quad \forall \psi \in \mathcal{H}^B
\end{equation}
\vspace{6ex}

The proof is simple---by writing out (31) in terms of $U_{{\rm b}i}$,
it follows from theorem 8.2 of \cite{nielsen} on the freedom of CP-map
decomposition that  
\begin{equation}
\sqrt{p'_{i}} U_{1i} = \sum_j \sqrt{p_j} V_{ji} U_{0j}
\end{equation}
for a unitary matrix {\bf V}.  This operator relation guarantees that the
state relation (32) is satisfied for all $| \psi \rangle$. 

When (31) is satisfied and (18) is used to compute the cheating $U^A (\psi )$ according to Theorem 2,
it is found to be independent of $| \psi \rangle$ and is given by the {\bf
V} of (33) due to the fact that the matrix $\langle \phi'_i | \phi_j \rangle$ becomes ${\bf V}$ multiplied by the inner product matrix $\langle \phi_i |
\phi_j \rangle$ which is nonnegative. Indeed, the {\bf V} in (33) is
also determined by following the usual impossibility proof for any
$\psi$. (Note
that the Schmidt decomposition plays no role in the proofs of 
theorems 2 and 3 and in the results used in their proofs. Indeed, 
Jozsa's proof of Ulhmann's theorem in \cite{jozsa}, which involves the
Schmidt decomposition, can also be simplified along the line in the
proof of Theorem 2 in Appendix A.)  Theorem 3 is significant in that
it shows it is operator, not state, entanglement that is needed in the
presence of state randomness.

Under the condition
\begin{equation}
\rho^{BC}_0 (\Psi) = \rho^{BC}_1(\Psi)
\end{equation}
for one fixed $| \Psi \rangle$ of the form (24), one obtains similar to the proof of Theorem 3 that
\begin{equation}
\rho^{B}_0(\psi) = \rho^B_1(\psi) \quad \forall \psi \in {\rm span} \{ | \psi_k \rangle \}.
\end{equation}
Equation (35) implies, in particular, that $\rho^B_0(\psi_k) =
\rho^B_1(\psi_k)$ for each $| \psi_k \rangle$ and a fixed cheating transformation is available as above.  The restriction on the validity of (35), and hence the possibility of Adam's successful cheating, to states in the subspace spanned by $\{ |\psi_k\rangle \}$ is indispensable as shown in the example of Appendix B.  We can summarize our two major criticisms of the impossibility proof. First and foremost, it is not properly formulated so that under (34) or
\begin{equation}
\rho^{BC}_0(\Psi) \approx \rho^{BC}_1(\Psi)
\end{equation}
for one fixed $| \Psi \rangle$ of (24), it may be Babe but not Adam
who can cheat, either because she may sent $| \psi \rangle \not \in
{\rm span} \{ | \psi_k \rangle \}$, or there is a $\lambda_k \approx
0$ for which $\rho^B_0(\psi_k) \not \approx \rho^B_1(\psi_k)$. Secondly, even assuming $|\Psi \rangle$ is formed by Babe, there is no proof that there is any cheating transformation that would work for all $| \psi_k \rangle$.

Another way to formulate
the problem at hand is to use CP-map or superoperator to characterize
the transition from $\psi$ to $\rho^B_{\rm b}$, similar to the proof
of Theorem 3. If two general CP-maps between operators on
${\mathcal{H}}_1$ and ${\mathcal{H}}_2$ are approximately equal in the
sense of (25) with $\psi \in {\mathcal{H}}_1$, the question is what
approximate relation would obtain between the positive operators in
their respective decompositions. This question is a complicated one
for application to our present problem, partly because when $\epsilon = \| \rho^B_0(\psi) - \rho^B_1(\psi)\|_1$ gets small, the security parameter $n$ grows unbounded and the resulting ${\cal H}^B$ and $\rho^B_b$ change profoundly.  An infinite-dimensional nonseparable Hilbert space formulation of the problem appears necessary at the beginning.  Until the question is settled in favor
of impossibility, there is no general impossibility proof for
protocols employing anonymous states even just in the simple fashion
of (22).

The QBC formulation in this section, while more general than that of
the impossibility proof which is a proper formulation only if the
randomness in the protocol are all in (1)-(2), is still quite limited in scope.  Indeed, the
protocols of the following sections IV and V already do not fit into
the present framework exactly.  There are many other ways to introduce
classical randomness in a protocol.  Even though they can be
represented quantum-mechanically, once measurements are made to pin
them down they would function just as in a classical protocol,
manifesting in the different ways the measurement results can be
utilized.  Just in the case of classical protocols, it does not appear
possible to characterize all QBC protocols to a useful extent that
something general can be said about the corresponding cheating
probabilities.  We will present elsewhere a general formulation of the
QBC problem.  It will be evident that the situation is far more
intricate than the impossibility proof formulation (1)-(2).

\newpage
\section{\hspace{.2in}Bit Commitment Scheme that Contradicts the
Impossibility Proof}

In this section, a protocol will be given that contradicts the quantitative claim of the impossibility
proof, (IP) of (9) or (IP') of (8), without yet being
unconditionally secure in the sense (US) of (8).  Its extensions
to unconditionally secure protocols will be given in the next section
V. An intuitive description on how the QBC scheme may be
developed is first provided to explain the underlying logic.

According to the impossibility proof formulation, there is a state
$|\Phi_{\rm b} \rangle$  of (1) - (2) shared by Adam and Babe.
The most general attack by Adam 
after $|\Phi_{0} \rangle$  is committed is to apply a local $U^A$ on
$\mathcal{H}^{A}$ and then make a
measurement on $\mathcal{H}^{A}$, or just to make a measurement on
$\mathcal{H}^{A}$ as in (4)-(5), and opens b = 1.  It is evident, from the way states in $\mathcal{H}^{B}$ can be
affected this way as given by (5), that
if $M = 1$ in (1) Adam cannot affect $\rho^{B}_{0} = | \phi \rangle
\langle \phi |$ in $\mathcal{H}^{B}$ at
all. Unconditional security is impossible in this case because $\bar{P}^{B}_{c} \sim \frac12$
implies $| \langle \phi | \phi' \rangle | \sim 1$ and thus
$\bar{P}^{A}_{c} \sim 1$ by simply announcing b = 1.  If one lets $| \langle \phi | \phi' \rangle | \not\sim 1$
then Babe can cheat by measurement and the protocol is not
concealing.  Our protocols are to be developed form the following
sequence of steps in general.  To be specific, qubits will be used in
this section.

To begin, let $| \phi \rangle$ and $| \phi' \rangle$ corresponding to
${\rm b} = 0,1$ be orthogonal so that
Adam cannot cheat.  To defeat Babe's cheating, Adam may send to Babe
the information qubit among many
random {\em decoy} states, named for example by their temporal order, and
announce the information qubit position when he opens.  To prevent
Adam from the obvious cheating of sending in both 
$| \phi \rangle$ and $| \phi' \rangle$ and opening accordingly, an {\em anonymous} state  $| \psi \rangle$ is first
sent by Babe, with Adam generating $|\phi \rangle = U_0 | \psi \rangle
, | \phi' \rangle = U_1 | \psi \rangle$ for $U_0 = I, U_1, = R(\theta, C)$ a rotation by an angle $\theta$ on
some great circle $C$ on the qubit Bloch-Poincare sphere. The rotation
can be applied by Adam without knowing $|\psi \rangle$ assuming, as usual, that
the orientations of all the qubit Bloch spheres are known to both Adam
and Babe.  Thus, $\langle \phi | \phi' \rangle = 0$ for $|\psi \rangle
\in C$ and $\theta =\pi$. It can be intuitively expected
that Babe cannot then determine ${\rm b}$ with $\bar{P}^{B}_{c} \sim
\frac12$  in the presence of
sufficiently many decoy states. It should also be clear that Babe
cannot improve his $\bar{P}^{B}_{c}$ by entanglement to $| \psi \rangle$, because she already
chooses a $| \psi \rangle$ that allows her to make perfect
discrimination if she 
knows which qubit is the one she sent, and so she has no need to change
$|\psi \rangle$ when she tries to cheat.  

How about Adam's new
possibilities of cheating at this stage? In all uses of anonymous
states, the other party can always try to determine the state by
measurement on the single copy.  It is characteristics of quantum
physics that the state cannot be determined and cannot be cloned
\cite{wootters}-\cite{yuen4} arbitrarily accurately, if it is drawn
from a nonorthogonal set of
states.  However, Adam has a significant probability of success in such
attempts, thereby such single use of qubit cannot yield an
unconditionally secure protocol---$\bar{P}^{B}_{c} \sim
\frac12$  and $\bar{P}^{A}_{c} \not\sim 1$ but not $\bar{P}^{A}_{c} \sim 0$.  More
precisely, with $n$ being the number of decoys states plus $| \psi \rangle$, one would have
\begin{equation}
\lim_n \bar{P}^{B}_{c} = \frac12, \;\;\;\; 0 < \lim_n \bar{P}^{A}_{c} <1
\end{equation} 
The protocol is thus concealing and quantitatively cheat-sensitive for Adam.
If Adam indeed cannot do better than cloning, the impossibility proof is contradicted with (37) and thus is incorrect
as a general proof.

A way to achieve (37), which has important practical significance, is for
Babe to make verifying measurements on all the qubits before Adam
opens.  She would choose the basis corresponding to $\{|\psi \rangle,
R(\pi, C) |\psi \rangle \}$    for all $n$
qubits.  Babe can evidently check whether Adam opens correctly in a
perfect fashion when he
identifies the qubit. It is intuitively clear, and will be explicitly
proved below, that the protocol is concealing. By entangling to the
qubit in state $|\psi \rangle$ in the form 
\begin{equation} 
\lambda_0 U_0 |\psi \rangle | e_0 \rangle + \lambda_1 U_1 | \psi \rangle | e_1
\rangle
\end{equation}
Adam can find out Babe's measurement result but he cannot change it for
cheating, as a matter of course---whatever operations and measurements
he performs cannot affect the result Babe already obtained. A precise treatment of the above
protocol QBCp3m, a preliminary (not yet unconditionally secure)
protocol with Babe's 
measurement before opening, is detailed presently, to be followed by the security proof.

\newpage
\addcontentsline{toc}{section}{\hspace{.45in}{Protocol QBCp3m}}
\noindent
{\em PROTOCOL} QBCp3m

\indent
(i) Babe sends Adam a state $|\psi \rangle$ known only to herself, randomly picked
from a fixed known great circle $C$ on the Bloch sphere of the qubit
${\cal{H}}^{B}_{2}$.

(ii) Adam modulates $|\psi \rangle$ by $U_0 = I$ or $U_1 = R(\pi , C)$, rotation of $|\psi \rangle$ to its orthogonal
state on $C$, for $b = 0$ or $b = 1$.  He then picks $n-1$ qubits with
states independently and randomly chosen among all possible ones, and
places the modulated qubit ${\cal{H}}^{B}_{2}$ randomly among them.  He sends the
$n$ resulting qubits to Babe, each named by its position in the qubit
sequence from $1$ to $n$.

(iii)  Babe measures $\{ |\psi \rangle, R (\pi, C) | \psi \rangle \}$
on each qubit.  Adam opens by revealing the
position of ${\cal{H}}^{B}_{2}$ and the bit value. Babe verifies by
checking her measurement result on ${\cal{H}}^{B}_{2}$. 

\vspace{3ex}
We first show that this QBCp3m is concealing.  For each possible $i$th
position for ${\cal{H}}^{B}_{2}$ in the qubit sequence sent back by
Adam, the state is of the form, in ${\mathcal{H}}^B$,
\vspace{2ex}
\begin{equation}
\begin{array}{c}| \phi_{1} \rangle \cdots \cdots {U_b | \psi \rangle}
\;\cdots | \phi_{n} \rangle \\ \quad i \end{array}
\end{equation}
where each $|\phi_{j} \rangle , j \in \{1,\cdots, n \}$ and $j\neq i$, is, say, one of the four BB84 states on $C$ randomly
and independently chosen. The index ``i'' underneath the state 
$U_{\rm b} |\psi \rangle$ in (39)
indicates that it occupies the $i$th position.  Thus, the state to Babe
is of the form, in ${\mathcal{H}}^B$
\vspace{2ex}
\begin{equation}
\begin{array}{c}\rho^B_{\rm b} = \frac{1}{n} \sum \frac{I}{2} \otimes \ldots \otimes
{\sigma_{\rm b}} \otimes \ldots \otimes \frac{I}{2}, \\ i \qquad
\qquad \quad \; i \quad \; \end{array}
\label{babe_do}
\end{equation}
with $\sigma_{\rm b} = U_{\rm b} \sigma U^\dagger_{\rm b}$ when Babe send a
state $\sigma$ to Adam without entanglement. Note that it is
sufficiently for Adam to choose among two orthogonal states instead of
all possible ones for each qubit, and for Babe to choose among four
BB84 states instead of all in a great circle. While it should be clear
that Babe gains nothing with entanglement,
that situation will be dealt with later. From (40), one can evaluate
$\bar{P}^{B}_{c}$ straightforwardly since $\rho^{B}_{0} -\rho^{B}_{1}$  is diagonal in the product basis that
diagonalized $\sigma_0 - \sigma_1$ on each qubit.  Let $n = 2\ell + 1$ and $\lambda_+ \leq
1$ be the positive eigenvalue of $\sigma_0 - \sigma_1$, it is shown in
Appendix C that
\vspace{2ex}
\begin{equation} \label{label.D} \label{eq36}
\bar{P}^{B}_{c} - \frac12 = \frac{\lambda_+}{2^n} 
\left( \begin{array}{c} 2 \ell \\ \ell \end{array} \right)
\end{equation}

\vspace{3ex}
\noindent The optimal probability (41) is obtained with $\lambda_+ = 1$ when the
above product basis is measured and b is set to be 0 or 1 according
to a majority rote on the positive and negative outcomes corresponding
to the eigenvectors $|\lambda_+ \rangle$ and $|\lambda _- \rangle$.
From the standard bounds on binomial coefficients,
\begin{equation} \label{label.E} \label{eq37}
\frac{1}{4 \sqrt{\ell}} 
< \bar{P}^{B}_{c} - \frac12 
< \frac{1}{2 \sqrt{\pi \ell}} 
\end{equation}

\vspace{3ex}
\noindent The optimal strategy is thus still concealing with $\lim_n
\bar{P}^{B}_{c} = \frac12$, but it is better than guessing at the
qubit sent and then measure and decide on it alone, which yields ${P}^{B}_{c} = \frac12 (1+ 1/n)$. 

To show that entanglement does not change the above situation in the
simplest possible way, we would merely give a detailed proof that concealing is not
affected by Babe's entanglement. When she entangles
${\cal{H}}^{B}_{2}$ to a ${\cal{H}}^{C}$ she
would attach ${\mathcal{H}}^{C}$ to one of the qubits sent back by Adam.  The
resulting density operator is the same independently of which
particular qubit position she attaches $\mathcal{H}^{C}$ to, from symmetry.  From
the triangle inequality for trace norm [18], the distance between the
resulting $\bar{P}^{B}_{\rm b}$ is bounded by 
\begin{equation} \label{label.E} \label{eq38}
n\parallel \rho^{B}_{0}- \rho^{B}_{1} \parallel_1 \;\; \leq  \;\; 2+\parallel
\bar{\rho}^{B}_{0}-\bar{\rho}^{B}_{1} \parallel_1
\end{equation}
where the term 2 is the maximum possible [23, App A] distance
$\parallel \rho_{0} - \rho_{1} \parallel_1$ for any states
$\rho_0$ and $\rho_1$, corresponding to the case where ${\mathcal{H}}^{C}$ is attached
correctly to ${\mathcal {H}}^{B}_{2}$.  The $\bar{\rho}^{B}_{\rm b}$ are
the same as (40) because the mismatched ${\mathcal{H}}^C$
state does not affect the trace distance as a consequence of
\begin{equation} \label{label.E} \label{eq40}
\parallel \left( \rho - \rho' \right) \otimes \sigma \parallel_1 =
\parallel \rho
- \rho' \parallel_1
\end{equation}
Equation (44) follows immediately from evaluating the left-hand side in the 
diagonal representation of $\left( \rho - \rho' \right) \otimes \sigma$.
 Thus, the protocol is still
concealing from (42) and (43).  Actually, it can be shown that the
optimal $\bar{P}^{B}_{c}$ of (41) without entanglement remains optimal with
entanglement.  Note that our proof shows that the protocol is
concealing for any
$|\psi \rangle \in {\mathcal{H}}^{B}_{2}$
even though we may impose restriction on $|\psi \rangle$ in the binding proof or for
ease of implementation.

Since Babe's verifying measurement can be perfectly made before Adam opens, a ``no-clone" argument can be developed for binding. Adam cannot find out what
measurement basis, not to mention $|\psi \rangle$, Babe used by entangling the qubits to ${\cal{H}}^A$---the state on ${\cal{H}}^A$ is obtained by tracing over ${\mathcal{H}}^B$ and is independent, not only of $|\psi \rangle$, 
but of the specific measurement basis Babe uses (or no
measurement from her at all). Thus, he can gain no information from Babe's measurement
to help him cheat in any way.  One way for Adam to cheat is by cloning, as it is the same whether one wants to get $\{ | \psi \rangle, |\psi \rangle \}$ or $\{ |\psi \rangle, U |\psi \rangle \}$ for a known $U$.
The optimal cloning performance is a fixed number $p_A < 1$
independent of $n$.
The optimal one-to-two clone has been worked out for a variety of
criteria and state sets.  In the present situation, the state set is $C$
or the four BB84 states.  If the cloning is described by $|\psi
\rangle \rightarrow | \psi_{a{\rm b}}\rangle$ over two
qubits with marginal states $\rho_a$ and $\rho_{\rm b}$, the criterion
here corresponds to  
\begin{equation} \label{label.C} \label{eq41}
F_c = \frac12 \langle\psi|\rho_{a}|\psi\rangle_{av}+ \frac12
\langle\psi|U_{1}^{\dagger}\rho_{\rm b} U_{1}|\psi\rangle_{av}
\end{equation}
with average over a uniform distribution on the state set from which $|\psi
\rangle$ is drawn.  It seems that the existing results \cite{keyl}-\cite{ariano}
almost cover this case exactly \cite{circle}. Now, it appears that
Adam cannot do better than this optimum by {\em any} action because if
he could, he should have succeeded in cloning better than the optimal
cloner, a contradiction, according to the following reasoning. He
would have, by an objective physical procedure, succeeded in producing
clones among $n$ qubits, where he could identify which ones are the
clones.  If Babe did not measure first, this would not be surprising
because the two copies are obtained on two different conditional (upon
Adam's measurement result) states for Babe.  The fact that Adam can
identify both means that he could not just spread $n-1$ qubit states
uniformly on $C$, one of which would be close to $U | \psi \rangle$,
but he wouldn't be able to tell which one. That he is not able to
identify both simulatneously does not alter the fact he has cloned.
Alternatively consider the following situation with the
cloning of one copy of $| \psi \rangle \otimes | \psi \rangle$ into
$\{ |\psi \rangle \otimes |\psi \rangle, |\psi \rangle \otimes | \psi
\rangle \}$, for a criterion as (41), with optimum $p_A < 1$.  If Babe
gets identical measurement results on two sets of $n$ qubits sent back
to her by Adam, each obtained by the same procedure as above, Adam
would have succeeded in cloning $| \psi \rangle \otimes | \psi
\rangle$ by carrying out the two different identification procedures
on the two $n$-qubit sets and applying the results to both sets.  To
ensure that Babe could have the identical measurement results almost
surely, consider the following Gedankenexperiment.  Babe sends a large
number $N$ of identical states $|\psi \rangle$ to Adam, who carries
out the same objective physical preparation (cheating) procedure on
her $N$ $n$-qubit sets.  Babe performs her measurement on each and
every set, obtaining, with probability exponentially close to 1, pairs
of identical results that total $N'$ sets with $N'/N$ close to 1 for
sufficiently large $N$.  Adam would then have, via the above separate
identification procedure on each pair, succeeded in cloning in almost
all of the original $N$ sets. Both the above single-set argument and
the present $N$-set argument are valid, but a complete formalization of the arguments will be given elsewhere.

Note that, in this protocol, Adam cannot cheat any better by generating decoy states other than $I/2$.  Thus we have covered all
possible actions by Adam and Babe, and can summarize the above results as

\vspace{3ex}
\noindent
{\em Theorem} 4:

In protocol QBCp3m, Babe's optimal cheating probability can be made
arbitrarily close to $\frac{1}{2}$ for large number of qubits $n$,
while Adam's optimal cheating probability remains fixed and not
arbitrarily close to $1$.

\vspace{3ex}
What would happen to Adam's EPR attack in the above scheme 
if Babe performs her verifying measurement after he opens?  
One may have the protocol ``QBC3'' in reference \cite{yuen3} in which Babe disregards 
the $n-1$ qubits not first sent by her.  It is simpler to consider the following 
variant more in line with the impossibility proof formulation. 
Let $|\psi\rangle \in S = \{ |1\rangle, |2\rangle, |3\rangle,
|4\rangle \}$ where $|1 \rangle$ and $|2 \rangle$ are the vertical
and horizontal states on $C$, and $|3 \rangle$ and $|4 \rangle$ are the two
orthogonal diagonal ones, so together they make up the four standard
BB84 states on $C$. Consider the case where each of the other $n-1$
qubits sent by Adam has to be in $S' = \{|1 \rangle, |2 \rangle \}$.  Adam modulates $| \psi
\rangle$ by $U_b$ and opens by identifying the ${\mathcal{H}}^{B}_{2}$ position and the
states of all the qubits.  Babe verifies by performing the corresponding
projection measurements.  Let $| \psi \rangle$ be purified as
\begin{equation} \label{label.B} \label{eq42}
\frac14 \begin{array}{c} 4 \\ \sum \\ {\ell =1} \end{array} | \ell \rangle
| f_{\ell} \rangle
\end{equation}
for $| \ell \rangle \in S \subset {\mathcal{H}}^{B}_{2}$ and orthonormal$| f_{\ell}
\rangle \in {\mathcal{H}}^{C}$.  Let $U_{1j}, j \in \{ 2, \cdots, n \}$ be the unitary operator that swaps
qubit position 1 and $j$ on ${\mathcal{H}}^{B} = {\mathcal{H}}^{B}_{2}
{\bigotimes^{n}_{j=2}} {\mathcal{H}}^{j}_{2}$. 
On ${\mathcal{H}}^A \otimes {\mathcal{H}}^B$, Adam can form the
entanglement by employing orthonormal $| e_i \rangle \in
{\mathcal{H}}^{A},i \in \{ 1, \cdots, n \cdot 4^{n-1} \}$, 
with uniform or whatever probabilities, using $U_{1j}$
and $S'$. In analogy with QBCp3m, we have a preliminary protocol QBC p3u which is
close to a usual one in which Adam can launch EPR attacks.

\addcontentsline{toc}{section}{\hspace{.45in}{\bf Protocol QBCp3u}}

\vspace{3ex}
\noindent
{\em PROTOCOL} QBCp3u

\indent
(i) Babe sends Adam a state $| \psi \rangle$ known only to herself,
randomly picked form the four BB84 states on a fixed great circle C of
the qubit ${\cal H}^B_2$.

(ii) Adam modulates  $| \psi \rangle$ by $U_0 = I$ or $U_1 = R(\pi,
C)$ for b = 0 or 1. He then picks $n-1$ qubits with states
independently and randomly from two orthogonal states known to Babe, places the modulated qubit ${\cal H}^B_2$ randomly among them, and sends
the $n$ qubits to Babe in a named order. 

(iii) Adam opens by revealing the state of all the qubits and
identifying  ${\cal H}^B_2$.  Babe verifies by checking the corresponding
projections.  

\vspace{3ex}
\noindent
This protocol is concealing exactly as in QBCp3m. As shown in section III, the impossiblity proof does not cover this 
protocol.  Indeed, assuming Adam opens perfectly for ${\rm b}=0$ as in the impossibility proof, it can be shown that he cannot then cheat with $\bar{P}^A_c \sim 1$.  The basic reason is that he can only identify correctly on the decoy states, for arbitrary $|\psi \rangle$, by not involving $|\psi \rangle$ in the entanglement of the decoy states.  However, he cannot then rotate $|\psi \rangle$ to its orthogonal complement on $C$.  The full security proof covering the situation in which Adam does not open perfectly is being developed.

\newpage

\section{\hspace{.2in}Unconditionally Secure Bit Commitment Schemes}

The QBC protocol in the previous section that invalidates the
impossibility proof can be extended to fully unconditionally secure
protocols as described in the following. This may be expected because
if Adam cannot cheat nearly perfectly on one qubit, his cheating
probability can be brought exponentially close to zero in a sequence of
independent qubits.  To extend the above protocols in this
manner, first consider the case where ${\mathcal{H}}^B_2$ in QBCp3m 
is replaced by ${\bar{\mathcal{H}}}^B= \otimes^m_{k=1}
{\mathcal{H}}^B_{2k}$.  Let Babe send Adam a sequence of $m$ qubits
\begin{equation}
|\psi \rangle =|\psi^1 \rangle \otimes \cdots \otimes | \psi^{j} \rangle \cdots
\otimes | \psi^m \rangle, \qquad j \in \{1, \cdots, m \}
\end{equation}
Each $| \psi^j \rangle$ is randomly and independently chosen from the
same fixed great circle $C$ for all the $m$ qubits, and named by its
sequence position $j$ within ${\bar{\mathcal{H}}}^{B}$.  Adam applies $U_b$ to each of these qubits and
then randomly places ${\bar{\mathcal{H}}}^{B}$ among a sequence of $N-1$ quantum spaces 
${\mathcal{H}}^{B}_{\ell}$, each a product of $m$ qubits, with states
on all the  $m \left( N-1 \right) $ qubits randomly
and independent chosen from a fixed great circle $C'$.
  The total sequence or  product state
\begin{equation}
| \chi_1 \rangle \cdots | \chi_{\ell} \rangle \cdots | \chi_N \rangle,
  \qquad \ell \in \{1, \cdots, N \}
\end{equation}
is re-named by the new position 
$\ell$ and sent back to Babe. Apart from the modulated state in
${\bar{\mathcal{H}}}^{B}$,
each of the other $N-1$ $| \chi_{\ell} \rangle$ in ${\mathcal{H}}^{B}_ {\ell}$ is a 
product of $m$ qubit states.  Each of the $N$ state spaces would be referred to 
as a {\em qumode}.  Similar to (35),   
Adam knows, but Babe does not,
which $| \chi_{\ell} \rangle$ is the modulated $| \psi \rangle$, and he opens 
by giving Babe this information, 
but he does not know what the $| \psi^j
\rangle$'s are.  Before Adam opens, Babe measures on every qumode the product qubit 
basis given by $\{ |\psi^j \rangle , R \left( \pi, C \right) |\psi^j \rangle \}$ 
across the $m$ qubits, which diagonalizes $\rho^B_0 - \rho^B_1$.  She 
optimally decides on ${\rm b}$ by the majority of the two patterns of $| \psi^j \rangle$ 
and $R \left( \pi, C \right) |\psi^j \rangle$, the other patterns occurring with 
equal probability.

To prove concealing, the following argument is used in lieu of evaluating 
directly the trace distance.  For any fixed $m$, let $N$ be chosen large enough 
that the number of times a particular pattern of $|\psi^j \rangle$ in (47)
shows up in Babe's 
measurement on a random qumode is at least $(N-1) \left( 2^{-m} - \delta \right)$ 
for a small $\delta >0$, where $(N-1)2^{-m}$ is the average.  This is possible with a 
probability exponentially close to 1 from the Chernov bound.  The situation then 
becomes the same as the qubit case of section IV, with $n$ replaced by $N \left( 
2^{-m} - \delta \right)$ for the upper bound in (42), which can then be set 
to any desired small level by further increasing $N$.  Babe's possible 
entanglement can be handled as in (43).  Thus, the protocol is concealing.  Adam's 
optimal cheating probability is given by $\bar{P}^A_c = p^m_A$, which fixes $m$ 
for given ${\bar{P}}^A_c < \epsilon$.  We summarize the results.

\vspace{3ex}
\addcontentsline{toc}{section}{\hspace{.45in}{\bf Protocol QBC3m1}}
\noindent
{\em PROTOCOL} QBC3m1

\indent
(i) Babe sends Adam a product state (47), each $|\psi^j \rangle$ named by its position
and independently and randomly chosen from a BB84 state set $S$ in $C$.

(i) Adam modulates each and all $|\psi^j \rangle$ by $U_0 = I$ or $U_1
= R(\pi, C)$, then independently and
randomly place the exact sequence among $N-1$ qumodes, each a product of 
$m$ qubits randomly distributed on $S$.  He
sends the $N$ qumodes to Babe in a named order.

(iii) Babe measures the $m \;\;\{ | \psi^j \rangle, R \left( \pi,C \right) |\psi^j \rangle \}$ 
on each of the $N$ qumodes.  Adam opens by announcing which qumode is the modulated 
$|\psi \rangle$ and the bit value.  Babe verifies by checking her measurement result.

\vspace{3ex}
\noindent
{\em Theorem} 5:

\indent
Protocol QBC3m1 is unconditionally secure.

\vspace{3ex}

Variations of the protocol can be easily created without affecting the unconditional 
security.  For example, consider the case where Babe sends (47) to Adam which he 
returns in $m$ segments of $N$ qubits each, the $j$th one containing exactly 
one $|\psi^j \rangle$ from (47). 
Babe can then make a uniform
measurement on each $N$-sequence, deciding whether each such
$N$-sequence corresponds to a 0 or 1 by a majority vote, and the
overall b by a majority vote on the $m$ outcomes.

To show that such a protocol is concealing, one may first take care of Babe's
entanglement possibility to ${\mathcal H}^C$ by, similar to (43),
\begin{equation}
\parallel \rho^B_0 - \rho^B_1 \parallel_1 \leq [1-p(N,\!m)] \cdot
2+p(N,\!m) \cdot \parallel \bar{\rho}^B_0 - \bar{\rho}^B_1 \parallel_1
\label{} 
\end{equation}
where $p(N,m)=(1- \frac{1}{N})^m$ is the probability 
that none of the $m$ attached entangled qubits in ${\mathcal H}^C$ matches 
the actual qubit position, which can be made arbitrarily close to 1 for 
any fixed $m$ by making $N$ large. Then one argues that independent qubit probability distributions 
are obtained because for optimal $P^B_c$ Babe should not entangle across the qubits in
(47) as that would create additional randomness for the individual qubit measurements 
she would make, the latter needed since she has a vanishingly small 
probability to locate her own qubits.  (Indeed, there is no point for her to correlate 
the $| \psi^j \rangle$ in the first place, as an involved classical probabilistic 
argument would show.) 
From the independence of $| \chi_\ell \rangle$ 
and the $| \psi^j \rangle$ positions in the $m$ $N$-sequences, the optimal 
decision Babe can make is to decide on 0 or 1 on each of the $N$-sequences as the
$m=1$ case, and then take a majority vote to decide on $b$. 
 Let $p$ be the probability
of Babe's correct decision in each $N$-sequence.  Then $p$ is given by
(41) and bounded as in (42) with $n$ replaced by $N$.  The overall
$\bar{P}^B_c = \sum^{{(m-1)}/2}_{k=0} (^m_{\:k} ) p^k(1-p)^{m-k}$
can be made, for any fixed $m$, arbitrarily close to 1/2 by making $p$
arbitrarily close to 1/2, i.e., with $N$ sufficiently large, because
this $\bar{P}^B_c$ is a continuous function of $p$.
The value of $m$ is determined from $p^m_A < \epsilon$ from
cloning.  With ${\bar{P}}^B_c - \frac{1}{2} < \epsilon$, the
unconditional security proof is completed for the following

\vspace{3ex}
\addcontentsline{toc}{section}{\hspace{.45in}{\bf Protocol QBC3m2}}
\noindent
{\em PROTOCOL} QBC3m2

\indent
(i) Babe sends Adam a sequence of $m$ qubits, each
$|\psi^j \rangle$ named by its position and independently and randomly
chosen from a great circle $C$.

(ii) Adam modulates each and all $|\psi^j \rangle$ by $U_0 = I$ and
$U_1 = R(\pi,C )$, then places each $U_b | \psi^j \rangle$
independently and randomly among the $j$th of $m$ succeeding
$N$-sequences of qubits, the states of all the other qubits independently
and randomly chosen. He sends the $n=mN$ succeeding qubits with their position names to Babe.

(iii)  Babe measures $\{|\psi^j \rangle, R(\pi, C)|\psi^j \rangle \}$
on the $N$ qubits of the $j$th sequence for all $j$.  Adam
opens by revealing the positions of $U_{b}|\psi^j \rangle$ and the
bit valve.  Babe verifies by checking her measurement results on these qubits.  

\vspace{3ex}
\noindent
{\em Theorem} 6:

\noindent
Protocol QBC3m2 is unconditionally secure.

\vspace{3ex}

Protocols QBC3u1 and QBC3u2 can be introduced similar to the last
section.  They are omitted here since their full security proofs are not yet available.

\newpage
\section{\hspace{.2in}Conclusion}

In this paper we have explicitly detailed two major ways in which the
QBC impossibility proof fails as a general proof. There are two
corresponding significant general issues concerning the impossibility proof.  One is
that classical randomness and the corresponding information flow
between the two parties may play a significant role in a general
protocol. Such a role has not been completely characterized for the
classical case, and cannot be simply eliminated by quantum
purification.  This points to the more general, second issue: how one
can characterize all possible QBC protocols at all when one has not been
able to do that for any type of classical cryptographic protocols.  In particular, 
there are many possible protocols with random numbers generated by Adam and 
Babe during various stages of a protocol, necessitating uniformity conditions similar to (25) 
that would intertwine in a complicated classical way that is not resolved by quantum purification. As things stand, it is even open whether a 
perfectly secure QBC protocol is possible, given the limited scope of Theorem 3.

In any event, it is possible to have unconditionally secure quantum
bit commitments, as protocols QBC3m1 and QBC3m2 demonstrate.  Equally
significantly, these protocols can be carried out without any quantum
memory to be used between commitment and opening.  In applications to
key management or identification/authentication, such required quantum
memory would be very long on microscopic scale, at least for network
type situations.  It is unrealistic to expect that such quantum memory
would become available in any reasonable amount of time.  Thus, these 
protocols represent 
a major step in advancing the possible practical use of quantum bit
commitment.  Moreover, each qubit in the protocol can be replaced by a
full optical field mode and qubit state by large-energy coherent
state, without affecting the essential underlying operations, thus
making the protocol even easier to implement. A full description
of such protocols and quantitative tradeoffs between security
and complexity will be given in a future paper.

\newpage
\addcontentsline{toc}{section}{Acknowledgment}

\section*{Acknowledgment}

I would like to thank M. D'Ariano and M. Ozawa for useful discussions.

This work was supported in part by the Defense Advanced Research
Project Agency and in part by the Army Research Office.

\newpage

\addcontentsline{toc}{section}{Appendix A: Proof of Theorem 2}

\renewcommand{\theequation}{A\arabic{equation}}
\setcounter{equation}{0}

\section*{Appendix A \\
Proof of Theorem 2}

By choosing $|e_i \rangle = | e^{'}_1 \rangle$ in (1)-(2), one obtains

\begin{equation}
|\langle \Phi_1 | U^A| \Phi_0 \rangle |= | tr {\bf U \Lambda} |
\end{equation}

\noindent
The maximum of $| tr {\bf U \Lambda} |$ over all unitary ${\bf U}$ is
attained when  ${\bf U \Lambda}$ is
nonnegative definite with maximum value given by $tr |{\bf \Lambda}|$
[17, p.43].  Thus  ${\bf U}$ is
determined by the polar decomposition (generalization to infinite
dimensional space can be obtained via maximal partial isometry) of ${\bf \Lambda}=|{\bf \Lambda}|{\bf U}^\dagger$.

With $p_i = p^{'}_i , {\bar{P}}^A_c$ is given by (20) and is thus
bounded above by $\sum_i |{\bf \Lambda} |_{ii}$ which is just $tr |
{\bf \Lambda} | = {F}$.  For a
set of probabilities $\alpha_i$ and complex numbers $\lambda_i$, one has
\begin{equation}
\sum_i \alpha_i | \lambda_i |^2 \geq | \sum_i \alpha_i \lambda_i |^2
\end{equation}
as a consequence of Jensen's inequality and the concavity of the
function $x \mapsto x^2$.  The lower bound of (21) follows from (A2)
with $\alpha_i = p_i$ and $\lambda_i = \sqrt{\tilde{p}_i / p_i} \langle
\tilde{\phi}_i | \phi^{'}_i \rangle$, valid for $p_i \neq p^{'}_i$.

\newpage
\addcontentsline{toc}{section}{Appendix B:  Example on Proper Concealing}

\renewcommand{\theequation}{B\arabic{equation}}
\setcounter{equation}{0}

\section*{Appendix B \\
Example on Proper Concealing}

In the notations of sections II-III, the following example shows that for (24), even with no small $\lambda_k$, the condition (34) does not imply that Adam can cheat as claimed by the impossibility proof.

Consider 2-qubit ${\cal H}^B = {\cal H}_2 \otimes {\cal H}_2$ and $|\Psi \rangle \in {\cal H}^B \otimes {\cal H}^C$ given by
\begin{equation}
| \Psi \rangle = \frac{1}{\sqrt{2}} (|a \rangle |a' \rangle |f_1 \rangle + |a' \rangle |a \rangle |f_2 \rangle )
\end{equation}
where $|a \rangle$ and $|a' \rangle$ are two openly known orthogonal
states in ${\cal H}_2$, and $|f_i \rangle$ are orthonormal in ${\cal
H}^C$, which is also a qubit. The operations are taken to be $p_1 =
p'_1 = p_2 = p'_2 = \frac{1}{2}$, $U_{01}=I$, $U_{02} = P$ the
permutation operator switching the two qubit positions in ${\cal
H}^B$, $U_{11}=R$ a rotation that brings $|a \rangle$ to $|a' \rangle$
and $|a' \rangle$ to $|a \rangle$, $U_{12} = RP$.  It follows easily
that, after entanglement by Adam, $\rho^{BC}_0(\Psi) =
\rho^{BC}_1(\Psi)$ and he can cheat perfectly when Babe forms (B1).

However, it is Babe who can actually cheat perfectly in this
situation.  Instead of sending (B1) she can send $|a \rangle |a
\rangle \in {\cal H}^B$ instead, which would defeat Adam's cheating
and allows herself to cheat.  The underlying reason is, of course,that
(31) or (25) is not satisfied, and $|a \rangle | a \rangle \not \in
{\rm span}\{ |a \rangle |a' \rangle, |a' \rangle |a \rangle \}$,
violating the condition required for (34)-(35).  Clearly, there is no
reason why Babe wants to be honest so Adam can cheat.  Thus, the
impossibility proof formulation, which does not have a condition such as (36), is not a meaningful one in the presence of random numbers, with consequent incorrect claim on same situation.

\newpage

\addcontentsline{toc}{section}{Appendix C:  Evaluation of Trace Distance}

\renewcommand{\theequation}{C\arabic{equation}}
\setcounter{equation}{0}

\section*{Appendix C \\
Evaluation of Trace Distance}

One straightforward way to evaluate $|| \rho^B_0 - \rho^B_1
\parallel_1$ for $\rho_{\rm b}$ of (40) is to directly compute the
trace norm in the product basis spanned by $\{ |\lambda + \rangle , |
\lambda - \rangle \}$ for each qubit.  Let $k$ be the number of
$|\lambda - \rangle$ in a product-basis vector.  One has, from a
direct counting calculation,
\begin{equation}
\parallel \rho^B_0 - \rho^B_1 \parallel_1 = \frac{\lambda_+}{n2^{n-1}}
\begin{array}{c} n \\ \sum \\ k=0 \end{array} 
\left( 
\begin{array}{c}
{n} \\ k 
\end{array} \right) | n-2k |
\end{equation}
The binomial sum in (C1) can be evaluated in closed form. With $n=2
\ell +1$,
\begin{equation}
\begin{array}{c} 
n \\ \sum \\ k=0 \end{array} 
\left( \begin{array}{c} 
n\\k
\end{array} \right) 
| n-2k |=2 (2 \ell +1) 
\left( \begin{array}{c} 
2 \ell \\ \ell 
\end{array} \right)
\end{equation}

\noindent
Equation (41) follows from (C1)-(C2).

\newpage
\addcontentsline{toc}{section}{Appendix D: Simple Summary of
Protocols QBCp3m, etc.}

\renewcommand{\theequation}{D\arabic{equation}}
\setcounter{equation}{0}

\section*{Appendix D \\
Simple Summary of Protocols QBCp3m, etc.}

The statement, underlying logic, and security of protocol QBCp3m  can be simply presented as follows. The detailed proofs are given
in the paper.

Let Babe send Adam a qubit in state $| \psi \rangle$ known only to herself, 
$|\psi \rangle \in C \subset {\mathcal{H}}^{B}_{2}$ in a fixed great circle
$C$ of the qubit Bloch sphere.  Depending on b = 0 or 1, Adam leaves it
alone or rotates it to its orthogonal state $| \psi' \rangle$, then
sends it back to Babe among a number $n-$1 of random decoy qubit states.
Independently of b, Babe can make the same qubit measurement of the
basis $\{ | \psi \rangle, | \psi' \rangle \}$ on every of the $n$
qubits before 
Adam opens.  The protocol is still concealing with $\bar{P}^B_c
\rightarrow \frac12$ as $n \rightarrow \infty$,
because she does not know which qubit is the one she sent.  It is
clear that Babe cannot determine b any better by sending $| \psi \rangle
\notin C$ or by entangling $|\psi \rangle$.  Because Adam cannot gain any information on Babe's 
measurement basis via entanglement, his optimal cheating probability $\bar{P}^A_c$ is given by
an appropriate one-to-two clone fidelity $p_A$, which is independent of $n$
and not arbitrarily close to 1.  As he has to open 0 and 1 on two
different qubits given Babe already measures, the optimality of $p_A$
would be contradicted if he can do any better.  Thus far, the
quantitative claim of the impossibility proof, (IP) of (9) or (IP$'$) of
(10), has been invalidated by the above protocol QBCp3m.  More
significantly, it shows that the impossibility proof
formulation misses a whole class of protocols in which Babe can make
the verifying measurement independently of b before Adam opens. 

It is straightforward to extend QBCp3m to unconditionally secure
protocols, such as QBC3m1 and QBC3m2, by having Babe send Adam a sequence of $m$ independent $|
\psi_i \rangle$'s with $p^m_A$ set to any arbitrarily small
value $\epsilon$.  Adam sends back each of the m uniformly modulated qubits
in different restricted ways among $n$ qubits.  Babe makes the corresponding measurements 
before Adam opens.  The resulting protocols are concealing with $n$
sufficiently large for any fixed $m$, which is determined by Adam's
optimal cheating probability $\bar{P}^A_c = p^m_A$, and are thus 
fully unconditionally secure in the sense (US) of (8).


\begin{thebibliography}{99}
\bibitem{adam1}Adam does not have to use the agreed upon $p_i$ and
$p'_i$ either.  However, it is to his advantage to do so in the
standard forumlation, because the protocol is then concealing but he
can still cheat.
\bibitem{circle}G. M. D'Ariano has been able to show that $F_c = \frac 
{1}{2} \left( 1+ \frac{1}{\sqrt{2}}\right)$ for the criterion (39) when $| \psi \rangle$ 
is drawn either from a great circle or from the four BB84 states on a great circle.
 (private communications).
\end{thebibliography}
\end{document}